\documentclass[preprint,showpacs,showkeys,preprintnumbers,amsmath,amssymb]{revtex4}

\usepackage{graphicx}
\usepackage{dcolumn}
\usepackage{bm}

\begin{document}

\preprint{}

\title{Noninformative Quantum $q$-Priors}

\author{Paul B. Slater}%
\email{slater@kitp.ucsb.edu}
\affiliation{%
ISBER, University of California, Santa Barbara, CA 93106\\
}%
\date{\today}

\begin{abstract}
We find, in an analysis involving four prior probabilities ($p$'s),
that the {\it information-theoretic}-based comparative 
noninformativity test devised by Clarke, and applied by Slater 
in a quantum setting, yields a ranking 
($p_{F_{q=1}} > p_{B} > p_{B_{q=1}trunc} >p_{F}$) {\it fully} 
consistent with Srednicki's 
recently-stated criterion for priors
of ``biasedness to
pure states''. Two of the
priors are formed
by {\it extending} certain metrics of quantum-theoretic interest
from three- to four-dimensions --- by incorporating 
the $q$-parameter
(nonextensivity/Tsallis index/escort parameter). 
The 
three-dimensional metrics are the 
Bures (minimal monotone) metric over the two-level quantum systems and 
the Fisher information metric over the corresponding 
family of Husimi distributions.
The priors $p_{B}$ and $p_{F}$ are the ({\it independent}-of-$q$) 
normalized volume elements of these metrics, 
and $p_{F_{q=1}}$ is the normalized volume element of the $q$-{\it extended} 
Fisher information metric, with $q$ set to 1. 
While 
originally intended  to similarly be
the $q$-extension of the Bures metric, with
$q$ then set to 1, the prior $p_{B_{q=1}trunc}$, actually
entails 
the {\it truncation} of the only 
{\it off}-diagonal entry of the extended Bures 
metric
tensor. Without this truncation, 
the $q$-extended Bures volume element is  {\it null}, as is also the case 
in two
other quantum scenarios we examine.

\end{abstract}


\maketitle
\section{Introduction}
Some fifteen years ago, Wootters asserted that ``there does not seem to be any natural measure on the set of all mixed states'' \cite[p. 1375]{wootters}. He did, however, consider random density matrices with all eigenvalues {\it fixed}.
He remarked that once ``the eigenvalues are fixed, then all the density 
matrices in the ensemble are related to each other by the unitary group, so
it is natural to use the unique unitarily invariant measure to define the
ensemble'' \cite[p. 1375]{wootters} (cf. \cite{mjwhall}).

Arguing somewhat similarly,
Srednicki recently proposed that
in choosing a prior distribution over density matrices, ``we can use the
principle of indifference, applied to the unitary symmetry of Hilbert space,
to reduce the problem to one of choosing a probability distribution for the
eigenvalues of $\rho$. There is, however, no compelling rationale for any
particular choice; in particular, we must decide how biased we are towards
pure states'' \cite[p. 6]{srednicki}.

\begin{figure}
\includegraphics{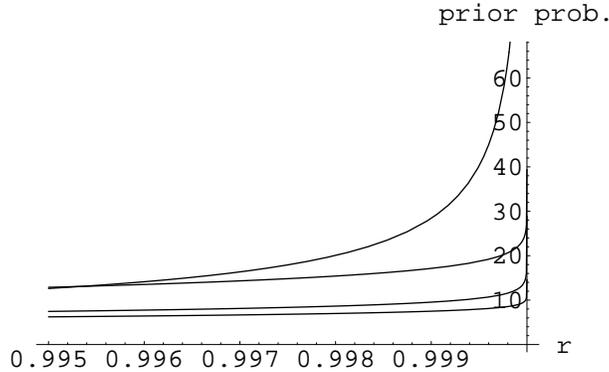}
\caption{\label{fig:biasedness}Four 
univariate marginal prior probability 
distributions in the near-to-pure-state region $r \in [1-\epsilon, 1]$, 
where $r$ is the radial coordinate in the Bloch sphere representation
of two-level quantum systems, and $r=1$ corresponds to a pure state. 
The order of dominance
{\it fully} complies with that (\ref{ordering}) obtained by
the information-theoretic-based 
comparative noninformativity test}
\end{figure}

In this study, we introduce evidence that Srednicki's 
standard of biasedness is, in effect,
incorporated into an information-theoretic (``comparative noninformativity'')
test --- originally devised by Clarke \cite{clarke} --- that has previously
been applied by Slater to quantum systems
\cite{compnoninform,slaterHusimi}.
We examine a
certain four prior 
probability distributions, denoted $p_{B_{q=1}trunc}, p_{F_{q=1}}, 
p_{B}$ and $p_{F}$. The first two are constructed in the
framework of {\it nonextensive statistical mechanics} \cite{AbeBagci}, 
being (at least, partial in the case of $p_{B_{q=1}trunc}$) 
$q$-extensions of the last two, which are normalized
volume elements of (classically and quantum) {\it monotone} metrics 
\cite{petzsudar}. 
The 
ranking in order of decreasing noninformativities that 
we obtain
\begin{equation} \label{ordering}
p_{F_{q=1}} > p_{B} > p_{B_{q=1}trunc} >p_{F}
\end{equation}
will prove (Fig.~\ref{fig:biasedness}) to be {\it fully} consistent with
the 
Srednicki ordering
according to biasedness to pure states.
\section{Bures Metric}
The Bures (minimal monotone) metric --- the volume element of which we 
normalize to obtain one ($p_{B}$) 
of the four prior probability distributions 
of 
principal
interest here --- yields the 
statistical distance between neighboring 
mixed quantum states ($\rho$) \cite{sam,uhlmann}.
It provides an embedding of 
the Fubini-Study metric \cite[sec. IV]{petzsudar}, which gives
the statistical distance between neighboring
pure quantum states ($|\psi \rangle$) (cf. \cite{majtey}).
H\"ubner gave an explicit formula for the Bures distance 
\cite[p. 240]{Hubner} (cf. \cite{luozhang}),
\begin{equation} \label{hub1}
d_{B}(\rho_{1},\rho_{2})^2 
= 2 -2 \mbox{tr} (\rho_{1}^{1/2} \rho_{2} \rho_{1}^{1/2})^{1/2}.
\end{equation}
Further, he expressed it 
in infinitesimal form as \cite[eq. (10)]{Hubner}
\begin{equation} \label{hub2}
d_{B}(\rho,\rho +d \rho)^2 =\Sigma_{ij} \frac{1}{2} \frac{|<i|d \rho| j>|^2}{\lambda_{i}+\lambda_{j}},
\end{equation}
where the $\lambda_{i}$'s are the eigenvalues  and the 
$|i \rangle$'s, the eigenvectors of $\rho$.
\subsection{Three-Dimensional Case}
In \cite{slaterBures}, using the familiar Bloch sphere (unit ball in Euclidean
3-space) representation of the two-level quantum systems ($2 \times 2$ 
density matrices),
\begin{equation} \label{nonescortDensityMatrix}
\rho = \frac{1}{2}  \left( \begin{array}{ccc}
1+z & x- i y \\
 x+ i y & 1-z\\
\end{array} \right), \hspace{.5in} r^2 = x^2+y^2+z^2 \leq 1,
\end{equation}
it was found (cf. \cite[p. 128]{mjwhall}), here converting from cartesian to spherical coordinates,
\begin{equation} \label{sphcoord}
x=r \cos{\theta_{1}}, \hspace{.3in} y=r \sin{\theta_{1}} \cos{\theta_{2}}, 
\hspace{.3in} z=r \sin{\theta_{1}}
\sin{\theta_{2}},
\end{equation}
that
\begin{equation} \label{BuresMetric}
d_{B}(\rho,\rho +d \rho)^2=\frac{1}{4} \Big(\frac{1}{(1-r^2)} dr^2 
+dn^2 \Big).
\end{equation}
The term $dr^2$ corresponds to the radial component of the metric
and $dn^2$, the tangential component ($dn^2= r^2 d \theta_{1}^2
+r^2 \sin^2{\theta_{2}}$).
In the setting of the quantum 
{\it monotone} metrics --- the 
Bures metric serving as the
{\it minimal} monotone one --- it is appropriate to express the tangential component of 
the Bures metric 
(\ref{BuresMetric}) in the form \cite[eq. (3.17)]{petzsudar},
\begin{equation} \label{unextendedBures}
\Big( (1+r) f_{B}(\frac{1-r}{1+r}) \Big)^{-1},
\end{equation}
where $f_{B}(t)= \frac{1+t}{2}$ is an {\it operator monotone} function 
\cite{lesniewski}.

The volume element of the Bures  metric 
(\ref{unextendedBures}) is $\frac{r^2 \sin{\theta_{1}}}{ 8 (1-r^2)}$,
which can be normalized to a {\it prior} probability distribution
over the Bloch sphere,
\begin{equation} \label{Buresprior}
p_{B}= \frac{r^2 \sin{\theta_{1}}}{\pi^2 (1-r^2)}.
\end{equation}
\subsection{Four-Dimensional Case}
Now, we can construct a {\it four}-dimensional family of 
(properly normalized/unit trace) $2 \times 2$ {\it escort} density
matrices (cf. \cite{naudts}),
\begin{equation} \label{escortDensityMatrix}
\rho_{\{q\}}= \Big( (1-r)^q+(1+r)^q \Big)^{-1}  \left( \begin{array}{ccc} 
1+z & x- i y \\ 
 x+ i y & 1-z\\
\end{array} \right)^q,
\end{equation}
for which $q=1$ recovers the standard Bloch sphere  representation 
(\ref{nonescortDensityMatrix}).
Applying H{\"u}bner's 
formula (\ref{hub2}), we have found that the {\it extended}
Bures metric (now incorporating the $q$-parameter)
has the form
\begin{equation} \label{extendedBures}
d_{Bures_q}(\rho,\rho+d \rho)^2= \frac{1}{4 (1+W^q)^2} \Big( 
W^q \log^2{W} dq^2 + \frac{4 q W^q \log{W}}{r^2-1} dq dr + 
\end{equation}
\begin{displaymath}
+ 4 \frac{q^2 W^q}{(r^2-1)^2 } 
dr^2
+ \frac{(-1+W^q)^2}{r^2 } dn^2 \Big),
\end{displaymath}
where $W=\frac{1-r}{1+r}$, that is, the ratio of the two 
eigenvalues of $\rho$.

\begin{figure}
\includegraphics{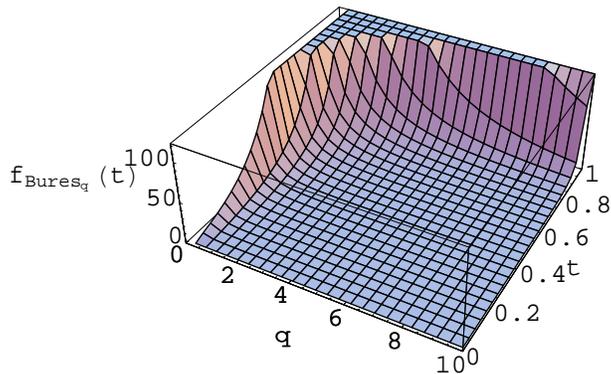}
\caption{\label{fig:monotonicgeneralization}The function $f_{Bures_{q}}(t)$
that yields the {\it tangential} component (\ref{generalf}) of the extended 
(four-dimensional) Bures metric (\ref{extendedBures})}
\end{figure}

The {\it tangential} component of the metric (\ref{extendedBures}) can
be expressed as $((1+r) f_{Bures_{q}}(W))^{-1}$, where
\begin{equation} \label{generalf}
f_{Bures_{q}}(t) =\frac{2 (1+t) (1+t^q)^2}{(-1+t^q)^2}.
\end{equation}
This 
bivariate function appears (Fig.~\ref{fig:monotonicgeneralization}) 
to be monotonically-increasing for any
fixed $q$ (cf. \cite{petzsudar}).

Now, in the earlier stage of our 
analyses, due to a programming oversight, we were under
the impression that the off-diagonal $dq dr$ term of (\ref{extendedBures})
was simply zero. If we do employ the fully correct form, 
with this $dq dr$ term included, we find that the
volume element is {\it null}. This, of course, could not yield a 
meaningful prior probability distribution. However, 
having proceeded under the impression that the $dq dr$ term was null,
we obtained a number of results that appear to be of interest 
and of some relevance.
Therefore, for much of this study, we will treat the $dq dr$ term as null, and
thus deal with a {\it truncated} $q$-extended Bures metric.

In the context of the harmonic oscillators states, 
Pennini and Plastino have argued that, in addition to the physical 
lower bound (ignorance-amount) of $q \geq 0$  that 
in a quantal regime,  
$q$ can be no 
{\it less}
than 1 \cite{pennini} --- due to the Lieb bound
on the Wehrl entropy \cite{lieb}. However, for the two-level quantum systems 
to the study of which we
restrict ourselves here, the lower  
bound on the Wehrl entropy is $\frac{1}{2}$  \cite[eq. (12)]{schupp}.
We, thus, consider $q \in [\frac{1}{2},\infty]$ to be the range of possible
values of the escort parameter $q$. In practice, though, we will, for numerical  and graphical 
purposes and normalization of the (divergent over $q \in [1/2,\infty]$) 
truncated extended Bures 
volume element (Sec.~\ref{qInference}), 
consider that $q \in [\frac{1}{2},500]$.

\begin{figure}
\includegraphics{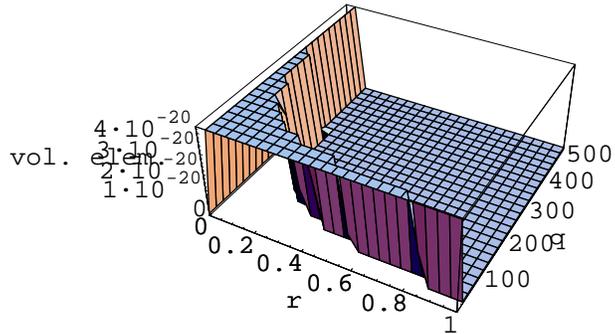}
\caption{\label{fig:BuresqVolElem}Two-dimensional marginal of the 
{\it truncated} four-dimensional extended Bures volume element (\ref{extendedBures})}
\end{figure}

\begin{figure}
\includegraphics{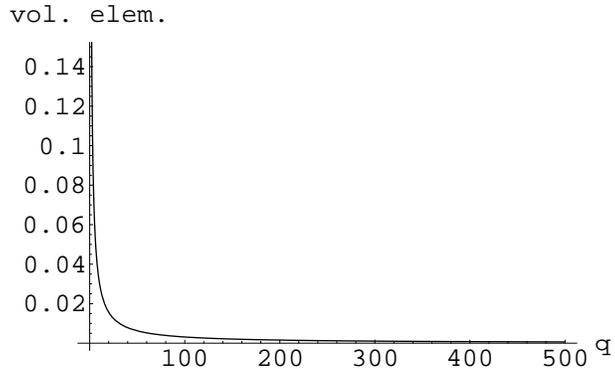}
\caption{\label{fig:BuresqVolElem1}One-dimensional marginal 
(\ref{exactprior}) over $q$ of the 
four-dimensional {\it truncated} 
extended Bures volume element (\ref{extendedBures})}
\end{figure}

In Fig.~\ref{fig:BuresqVolElem} we show the {\it two}-dimensional 
{\it marginal}
volume element of (\ref{extendedBures}) (after omission of the
$dq dr$ term)  --- integrating out the spherical angles, $\theta_{1}, \theta_{2}$, and leaving the radial coordinate $r$ and 
the escort parameter $q$.
In Fig.~\ref{fig:BuresqVolElem1}, further integrating out $r$, 
we show the corresponding {\it one}-dimensional marginal
volume element of (\ref{extendedBures}) (after omission of the $dq dr$ 
term) over $q$.
This (Fig.~\ref{fig:BuresqVolElem1}) has the exact expression 
\begin{equation} \label{exactprior}
\frac{\pi (1+\log{4})}{24 q}.
\end{equation}
This prior, thus, conforms to Jeffreys' rule --- as opposed to the
Bayes-Laplace rule, which would give a {\it constant} prior \cite{slaterlavenda}.

\begin{figure}
\includegraphics{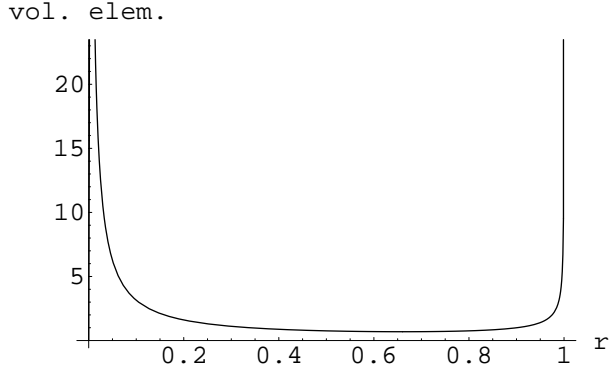}
\caption{\label{fig:BuresrVolElem1}One-dimensional marginal (obtained from 
(\ref{exactprior2})) over $r$ of the
four-dimensional extended Bures volume element (\ref{extendedBures}) after
omission of the off-diagonal $dq dr$ term}
\end{figure}

In Fig.~\ref{fig:BuresrVolElem1}, we integrate out $q \in [\frac{1}{2},
500]$,
leaving a (deep bowl-shaped)
 one-dimensional marginal over $r \in [0,1]$. (The corresponding 
marginal in the unextended Bures case is $\frac{\pi r^2}{2 (1-r^2)}$, 
so it is simply increasing with $r$, in that case.)
The associated  indefinite integral over $q$ is
\begin{equation} \label{exactprior2}
\frac{\pi \,\left( q\,W^q\,\left( 3 + W^{2\,q} \right) \,
       \log (W) - \left( 1 + W^q \right) \,
       \left( 2\,W^q + {\left( 1 + W^q \right) }^2\,
          \log (1 + W^q) \right)  \right) }{6\,
    \left( -1 + r^2 \right) \,
    {\left( 1 + W^q \right) }^3\,\log (W)}.
\end{equation}
(So, we obtain the function plotted in  
Fig.~\ref{fig:BuresrVolElem1} by substituting $q=500$ and $q=\frac{1}{2}$
into (\ref{exactprior2}) and taking the difference.)

For $q=1$, the {\it extended} Bures metric 
(\ref{extendedBures}) reduces to
\begin{equation} \label{extendedq1Bures}
ds_{Bures_{q=1}}(\rho,\rho+d \rho)^2= \frac{1}{16} (1-r^2) \log^2{W} dq^2 
- \frac{1}{4} \log{W} dq dr 
+ ds_{B}(\rho,\rho+d \rho)^2.
\end{equation}
Normalizing the volume element of this metric --- but 
first nullifying the off-diagonal $dq dr$ term --- to a (non-null) 
prior 
probability distribution over the Bloch
sphere, we obtain (cf. (\ref{Buresprior})),
\begin{equation}
p_{B_{q=1}trunc} =\frac{3}{4} \frac{r^2 \sin{\theta_{1}} \log{\frac{1}{W}}}{\pi (1 +\log{4})},
\end{equation}
one of the four priors that we rank (Fig.~\ref{fig:biasedness} and (\ref{ordering})) both by the comparative 
noninformativity test and Srednicki's biasedness criterion.

\subsection{Comparative Noninformativities in the Bures Setting}
The {\it relative entropy} (Kullback-Leibler 
distance/information gain \cite{lisa,vedralRMP}) 
of $p_{B}$ with respect to 
$p_{B_{q=1}trunc}$ [which we 
denote $S_{KL}(p_{B},p_{B_{q=1}trunc})$] --- that is,
the {\it expected} value with respect to $p_{B}$
of $\log{\frac{p_{B}}{p_{B_{q=1}trunc}}}$ --- is
0.101846 ``nats'' of information. Now, reversing arguments, 
$S_{KL}(p_{B_{q=1}trunc},p_{B}) =0.0661775$. (We 
use the {\it natural} logarithm, and
not 2 as a
base, with one nat equalling 0.531 bits.)
Let us convert --- using Bayes' rule --- these 
two (prior) probability distributions to {\it posterior}
probability distributions ($post_{B}$ and $post_{Bures_{q=1}}$), 
by assuming {\it three}  
pairs of spin measurements, 
{\it one}  each
in the x-, y- and z-direction, each pair yielding one ``up'' and one ``down''.
This gives us the {\it likelihood} function (cf. \cite[eq. (9)]{srednicki} 
\cite[eq. (4.2)]{bagan}),
\begin{equation} \label{lik}
L(x,y,z)= \frac{(1-x^2) (1-y^2) (1-z^2)}{64}
\end{equation}
(which we convert to the spherical coordinates (\ref{sphcoord}) 
in which we perform our Mathematica computations).

Then, we have $S_{KL}(post_{B}||p_{B_{q=1}trunc})= 0.169782$ and 
$S_{KL}(post_{Bures_{q=1}}||p_{B}) = 0.197657$.
The relative magnitudes of the information gains obtained by passing from priors to 
posteriors (0.101846 to 0.169782 and 0.0661775 to 0.197657) seems to suggest 
that $p_{B}$ is  
somewhat {\it more} noniformative than
$p_{B_{q=1}trunc}$. This is 
confirmed, using the 
testing structure given in 
\cite{compnoninform,slaterHusimi} (cf. \cite{srednicki}),
if we {\it formally} use a likelihood ($L(x,y,z)^{\frac{1}{2}}$), 
which is the square root of (\ref{lik}), to compute $post_{B}$ and $post_{Bures_{q=1}}$.
Then, we see a {\it decrease} in relative entropy from 0.101846 to
0.093849 and an {\it increase} from 0.0661775 to 0.114669. So, $p_{B}$
can be made {\it closer} to $p_{B_{q=1}trunc}$ by {\it adding} information
to it, but not {\it vice versa}, leading us to conclude that
$p_{B}$ is {\it more} noninformative than $p_{B_{q=1}trunc}$, 
since it assumes {\it less} about the data. 
(Let us note, 
however, that in the class of monotone metrics \cite{petzsudar}, the 
Bures or minimal monotone metric appears to be the {\it least}
noninformative (cf. \cite[sec. 5]{mjwhall}). The {\it maximal} monotone metric, on the other 
hand,  is {\it not}
normalizable to a proper prior probability distribution over 
the Bloch sphere \cite{compnoninform}. So, there is an interesting question of 
whether there exists a 
{\it single}, distinguished {\it normalizable} monotone metric which is
{\it maximally} noninformative.)
\section{Fisher Information Metric of Husimi Distributions}
Let us now move to a classical context, employing the 
(generalized) Husimi distributions 
\cite{monge}, 
rather than density matrices to represent the two-level quantum 
systems. 
Use of the Fisher information 
(monotone) metric \cite{chentsov,pap} is 
now indicated. To generate the (properly normalized) 
{\it escort} Husimi distributions 
($H_{\{q\}}$) (cf. \cite{pennini}), from 
the Husimi distribution ($H = H_{\{1\}}$), we employ 
the formula (cf. (\ref{escortDensityMatrix})),
\begin{equation}
H_{\{q\}} =2 \left( r + q\,r \right) \Big( -{\left( 1 - r \right) }^{1 + q} +
    {\left( 1 + r \right) }^{1 + q} \Big)^{-1} H^{q}.
\end{equation}

The tangential components of the Fisher information metric 
for the escort Husimi distributions 
($H_{\{q\}}$) are of
the form \newline 
$((1+r) f_{F_{q}}(t))^{-1}$, where \cite[eq. (29)]{slaterHusimi}
\begin{equation} \label{nonintegral}
f_{F_{q}}(t) = \frac{\left( -1 + q \right) \,{\left( -1 + t \right) }^2\,
    \left( -1 + t^{1 + q} \right) }{q\,
    \left( 1 + t \right) \,
    \left( 1 - q + t + q\,t - t^q - q\,t^q - t^{1 + q} +
      q\,t^{1 + q} \right) }.
\end{equation}
In \cite[sec. V.D]{slaterHusimi}, 
we succeeded in finding similarly
general (for all $q$) 
formulas for the denominators, but not the numerators, 
of the radial components.

\begin{figure}
\includegraphics{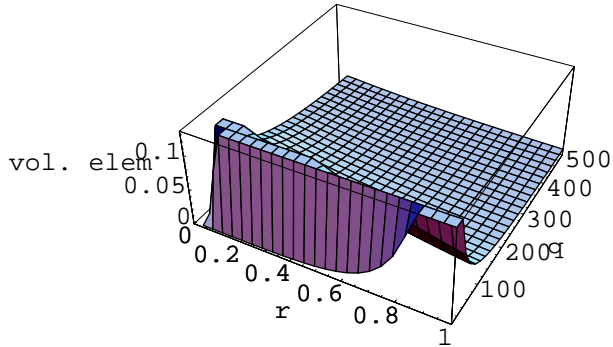}
\caption{\label{fig:HusimiqVolElem}Two-dimensional 
marginal of the four-dimensional extended 
Husimi volume element (\ref{extendedHusimi})}
\end{figure}

\begin{figure}
\includegraphics{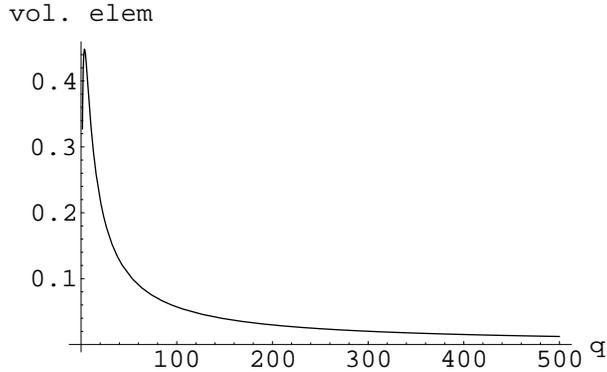}
\caption{\label{fig:HusimiqVolElem1}One-dimensional marginal over $q$ of the
four-dimensional extended Husimi volume element (\ref{extendedHusimi}). 
There is a peak near $q=3.59782$}
\end{figure}

\begin{figure}
\includegraphics{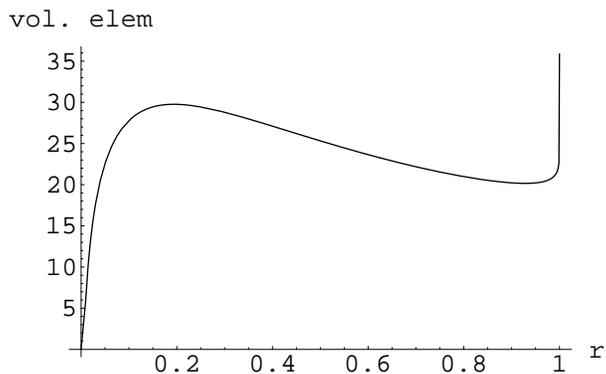}
\caption{\label{fig:HusimirVolElem1}One-dimensional marginal over $r$ of the
four-dimensional extended Husimi volume element (\ref{extendedHusimi}).
The upturn near $r=1$ may be due to (hard-to-avoid) numerical inaccuracy.}
\end{figure}

In Fig.~\ref{fig:HusimiqVolElem} we show (having to resort to some {\it 
numerical} 
integrations, since 
we lack explicit [$q$-general] expressions for certain of the metric 
elements) 
the counterpart to Fig.~\ref{fig:BuresqVolElem} 
for the four-dimensional extended {\it Husimi} metric.
Continuing with our numerical methods, we obtain the 
interesting unimodal curve (Fig.~\ref{fig:HusimiqVolElem1}) --- the peak being near $q=3.59782$, with a value there of 0.448488. This 
portrays the {\ one}-dimensional marginal Husimi volume element over $q$ (cf. Fig.~\ref{fig:BuresqVolElem1}).
In Fig.~\ref{fig:HusimirVolElem1} we show the 
(quite difficult-to-compute) one-dimensional marginal
over $r$ (cf. Fig.~\ref{fig:BuresrVolElem1}). (It appears the upturn 
near $r=1$ may be 
simply a numerical artifact. The difficulty consists in that, in some sense, we have to repeatedly perform  numerical integrations using results of other numerical integrations. It would be of interest to see how the curve changes as
the range of $q \in [\frac{1}{2},500]$ is modified.)
\subsection{Three-dimensional metric}

For the case $q=1$, the (unextended) three-dimensional 
Fisher information metric 
over the family of Husimi distributions takes the 
form \cite[eq. (2)]{slaterHusimi}
\begin{equation} \label{expressionHus}
ds_{F}(\rho,\rho+ d \rho)^2 = \frac{-2 r - \log (\frac{1-r}{1+r})}{2 r^3} dr^2 +
\Big((1+r) f_{F}(\frac{1-r}{1+r})\Big)^{-1} dn^2.
\end{equation}
Here,
\begin{equation} \label{zzz}
f_{F}(t)= \frac{(t-1)^3}{t^2-2 t \log{t}-1},
\end{equation}
which is the limiting case ($q \to 1$) of (\ref{generalf}).
To normalize the  volume element of this metric 
(\ref{expressionHus}) to a prior probability
distribution ($p_{F}$), we divide it by 1.39350989 \cite{slaterHusimi}.
\subsection{Four-dimensional metric}
In the extended (four-dimensional) 
case (cf. (\ref{extendedq1Bures})), {\it after} having set $q=1$,
we have,
\begin{equation} \label{extendedHusimi}
ds_{F_{q=1}}(\rho,\rho+ d \rho)^2= 
\Big( \frac{1}{4} -\frac{(-1+r^2)^2 \log^2{W}}{16 r^2} \Big) 
dq^2 
\end{equation}
\begin{displaymath}
+ \frac{2 r - (-1+r^2) \log{W}}{2 r^2} dq dr +ds_{F}(\rho,\rho+d \rho)^2.
\end{displaymath}
(So, the metric tensor here, in the same manner as in
the {\it untruncated} extended Bures case (\ref{extendedBures}), is not fully 
diagonal. We do {\it not} truncate the $q$-extended Fisher information
metric (\ref{extendedHusimi}) in any of our analyses.)
To normalize its (non-null) volume element 
to a prior probability distribution 
($p_{F_{q=1}}$) over the Bloch sphere, we must divide by 
0.24559293.
\section{Comparative Noninformativity Analysis}
We have that 
$S_{KL}(p_{F}||p_{F_{q=1}}) = 0.229666$
and $S_{KL}(p_{F_{q=1}}||p_{F}) = 0.170145$.
Further, using the likelihood (\ref{lik}), based on six hypothetical
measurements to generate posteriors, 
we obtain $S_{KL}(post_{F},p_{F_{q=1}}) = 0.70766$ and
$S_{KL}(post_{F_{q=1}}||p_{F}) =0.0641738$.
So, the comparative noninformativity test, which was initially developed by
Clarke \cite{clarke}, leads us to a firm conclusion that 
the four-dimensional-based probability distribution $p_{F_{q=1}}$
is {\it more} noninformative in nature than the three-dimensional-based
$p_{F}$.

Additionally, $S_{KL}(p_{B}||p_{F_{q=1}})=0.148269$ and 
$S_{KL}(p_{F_{q=1}}||p_{B}) = 0.0989669$. These are converted, 
respectively, to 0.283218 and 0.0842879 if we replace  the first arguments
of the two relative entropy functionals by posterior distributions based on
the (formal) square root ($L(x,y,z)^{\frac{1}{2}}$) of the likelihood function (\ref{lik}).
Thus, we can conclude that $p_{F_{q=1}}$ is also 
{\it more}
noninformative than $p_{B}$.

Further, $S_{KL}(p_{B_{q=1}trunc}||p_{F_{q=1}})=0.105463$
and $S_{KL}(p_{F_{q=1}}||p_{B_{q=1}trunc})=0.0914175$. 
Again, using the formal square root ($L(x,y,z)^{\frac{1}{2}}$) 
of the likelihood, we obtain
changes, respectively, to 0.245602 and 0.0408236. 
So, our
conclusion here is that $p_{F_{q=1}}$ is also more noninformative
than $p_{B_{q=1}trunc}$.
We already know from \cite{slaterHusimi} that $p_{B}$ 
is considerably more noninformative than $p_{F}$.

Continuing along these lines, 
$S_{KL}(p_{B_{q=1}trunc}||p_{F})=0.0191948$ and 
$S_{KL}(p_{F}||p_{B_{q=1}trunc})= 0.0234599$ (so the two 
distributions are relatively
close to one another).
Using  ($L(x,y,z)^{\frac{1}{2}}$)
to generate posterior distributions, the first statistic is altered 
(slightly decreased)
to 0.0143147, while the second statistic jumps to 0.1047772.

So, assembling these several relative entropy statistics,
we have the previously indicated ordering of the four priors (\ref{ordering}).
(The conclusions of the comparative noninformativity test appear to be
{\it transitive} in nature, although I can cite no explicit theorem to that 
effect.)
\subsection{Relation to Srednicki's Criterion for Priors}
In Fig.~\ref{fig:biasedness}, we show the one-dimensional
marginal probabilities of the four prior probabilities  over the radial 
coordinate $r$ in the near-to-pure-state range $r \in [.995,1]$.
The dominance ordering in this plot {\it fully} complies with
that (\ref{ordering}) found by the 
information-theoretic-based comparative noninformativity test. 
(We note that this ordering is {\it not} simply {\it reversed} near to the 
fully mixed state [$r=0$].)
Conjecturally,  
this could be seen as a specific case of some (yet unproven)
theorem --- perhaps utilizing the convexity and 
decreasing-under-positive-mappings 
properties \cite[p. 35]{ohya} of the relative entropy functional.

 So, the  information-theoretic (comparative-noninformativity)
test appears to incorporate Srednicki's criterion of
``biasedness to pure states'' \cite{srednicki}.
(Of course, it would be interesting to test the consistency between the 
comparative noninformativity test and Srednicki's criterion with a larger 
number of priors, as well as in higher-dimensional quantum settings 
(cf. \cite{slaterSPIN}).)
Srednicki does not explicitly 
observe that increasing biasedness to pure 
states corresponds to increasing noninformativity. He asserts
that ``we must decide how biased we are towards pure states''.

Srednicki focused on {\it two} possible priors. One was the uniform distribution over the Bloch sphere (unit ball). In \cite[sec. 2.2]{compnoninform},
we had concluded that this distribution was {\it less} noninformative than
$p_{B}$, in full agreement with contemporaneous work of 
Hall \cite{mjwhall}.
The second prior (``the Feynman measure''), which Srednicki points out is less biased to the pure states than the uniform distribution, 
was discussed in \cite{slaterlmp}. Neither of the two priors analyzed by
Srednicki corresponds to the normalized volume element of a {\it monotone} 
metric 
\cite{compnoninform,slaterlmp}.
\section{$q$-Extended Inference} \label{qInference}
In the setting of the $q$-parameterized escort density matrices
(\ref{escortDensityMatrix}),
the factor $\frac{1-z^2}{4}$ in the likelihood (\ref{lik}), giving the probability
(in the standard three-dimensional Bloch sphere setting) 
of one spin-up and one spin-down being measured 
in the $z$-direction, would be 
{\it replaced} by 
\begin{equation} \label{extendedlikelihood}
L_{q}(z) = \frac{r^2 (1+W^q)^2 -(-1+W^q)^2 z^2}{4 r^2 (1+W^q)^2},
\end{equation}
and similarly for the $x$- and $y$-directions. 
(For $q=1$, we recover $\frac{1-z^2}{4}$.)

It would be interesting to ascertain if the 
volume elements of the extended four-dimensional 
(truncated) Bures 
and
Husimi metrics ((\ref{extendedBures}) and (\ref{extendedHusimi})) could be integrated over the product of the Bloch sphere
{\it and} $q \in [\frac{1}{2},\infty]$ and normalized to (prior) probability distributions.
Then, using likelihoods 
incorporating the form (\ref{extendedlikelihood}), one 
could conduct the comparative noninformativity test in a {\it four}-dimensional
setting, rather than only the {\it three}-dimensional one employed 
throughout this study.
It turns out, however, that the three-fold 
integral --- holding $q$ fixed --- of the truncated 
volume element of 
(\ref{extendedBures}) over the
Bloch sphere is given by our formula (\ref{exactprior}).
Therefore, the four-fold integral of the one-dimensional marginal 
over the indicated product region with 
$q \in [\frac{1}{2},\infty]$ must
{\it diverge}. So, to achieve a {\it proper} probability distribution one
would have to truncate $q$ above a certain value.

Continuing along these lines, we omitted 
$q$ above 500 (and below $q=\frac{1}{2}$) 
and normalized 
the volume element of the 
(truncated) 
extended Bures metric to a proper probability
distribution. Then, the information gain 
with respect to such a prior, using $L_{q}(z)$, 
is
0.0597923  nats of information, while a
 {\it single} up or down measurement yields 0.134651 nats, and  two
measurements along the same axis 
giving the same outcome leads to an information gain of
0.349601. The analogous three (slightly {\it larger}) 
statistics, working in the unextended
framework (where $q$ does not explicitly enter, and is implicitly
understood to equal 1), using $p_{B}$ as prior, 
are, respectively,  $\frac{7}{6}-\log{3} \approx 0.0680544$, and
\begin{equation}
\frac{8\, _{p}F_{q}(\{ \frac{1}{2},1,
         2\} ,\{ \frac{3}{2},\frac{5}{2}\} ,1) -
      \pi \,\left( -5 + \log (64) \right) -6 -12\, K   }{6\,
    \pi } \approx 0.140186,
\end{equation}
(where $_{p}F_{q}$ denotes a generalized hypergeometric function and 
 $K \approx 0.915965594177$ is Catalan's constant) 
and $ \frac{59}{30} -\log{5} \approx 0.357229$. (We encountered 
numerical difficulties using Mathematica in attempting to extend these analyses
to measurements conducted in more than one direction, unless we restricted
$q$ to a range no larger than on the order of 10.)

One might also consider 
the possible relevance of 
$q$-analogs of the Clarke comparative noninformativity 
test, using $q$-relative entropy (Kullback-Leibler) divergence
\cite{johal,hirokisuyari}.

\section{Four-Dimensional $3 \times 3$ Density Matrices}
In \cite{slaterSPIN}, we considered an extension of the 
$2 \times 2$ density matrices (\ref{nonescortDensityMatrix}) to the 
$3 \times 3$ form (by incorporating an additional parameter $v$)
\begin{equation} \label{threebythreeDensityMatrix}
\rho = \frac{1}{2}  \left( \begin{array}{ccc}
v+z & 0 &  x- i y \\
0 & 2 -2 v & 0 \\
 x+ i y & 0 & v-z\\
\end{array} \right), \hspace{.5in} r^2 = x^2+y^2+z^2  \leq v^2; 
\hspace{.3in} 0 \leq v \leq 1,
\end{equation}
The Bures metric was found there to take the form
\begin{equation} \label{spin1extension}
d_{B_{n=3}}(\rho,\rho+ d \rho)^2 = 
\frac{1}{4} \Big( \frac{r^2-v}{ (1-v) (r^2-v^2)} dv^2 
+\frac{r}{r^2-v^2} dv dr +
+\frac{v}{v^2-r^2} dr^2 +\frac{1}{v} dn^2 \Big).
\end{equation}
(So, the tangential component is {\it independent} of $r$, as with 
(\ref{BuresMetric}) (cf. \cite{mjwhall}).)
Normalizing the volume element of (\ref{spin1extension}), 
we obtain the prior probability distribution 
\cite[eq. (18)]{slaterSPIN}
\begin{equation}
p_{B_{n=3}}
= \frac{3 r^2 \sin{\theta_{1}}}{4 \pi^2 v \sqrt{1-v} \sqrt{v^2-r^2}}.
\end{equation}
We have calculated that the (five-dimensional) $q$-extension of this metric
has a tangential component of the form
\begin{equation}
\frac{{\left( {\left( -r + v \right) }^q - 
       {\left( r + v \right) }^q \right) }^2}{4\,r^2\,
    \left( {\left( -r + v \right) }^q + 
      {\left( r + v \right) }^q \right) \,
    \left( {\left( 2 - 2\,v \right) }^q + 
      {\left( -r + v \right) }^q + 
      {\left( r + v \right) }^q \right) },
\end{equation}
but have not yet been able to derive simple forms for the other entries
of this metric tensor.

Numerical tests appear to indicate that the volume element
of this $q$-extended Bures metric tensor is (also) identically zero.
\section{$q$-Extension of the Bures Metric for the Abe-Rajagopal States}

Since our two attempts aove  to extend the Bures metric from an 
$n$-dimensional 
setting to an $(n+1)$-dimensional 
framework, by embedding the $q$ order parameter,
have 
yielded metrics (one of them being (\ref{extendedBures})) 
with zero volume elements, we were curious as to
whether or not we could obtain, in some other quantum context, a {\it 
nondegenerate}
$q$-extension of the Bures metric. In this regard, we 
turned our attention to the paper,
``Quantum entanglement inferred by the principle of maximum nonadditive
entropy'' of Abe and Rajagopal \cite{aberajagopal} (cf. \cite[eq. (14)]{tlb}).

Their principal object of study is a $4 \times 4$ density matrix
\cite[eq. (32)]{aberajagopal}, being ostensibly parameterized by
{\it three} variables, the order (nonadditivity) parameter $q$, 
the $q$-expected value $b_{q}$ of the Bell-CHSH observable and 
its dispersion $\sigma_{q}^2$. ({\it Two} of the four eigenvalues 
of the density matrix are always equal.
In \cite{slaterEuro}, it was asserted that
for the cases $q=\frac{1}{2}$ and 1, the associated {\it separability
probabilities} were equal to the ``silver mean'', that is, 
$\sqrt{2}-1 \approx 0.414214$ 
(cf. \cite{slaterJGP,slaterPRA}). We have confirmed these two probabilities
here --- at least in a numerical sense --- and also found that the Bures
volume of separable and nonseparable states 
is approximately  0.785398 [which we believe is an approximation
to $\frac{\pi}{4} \approx 0.7853981634$] 
for {\it both} $q=\frac{1}{2}$ and 1 
[as well as for $q=\frac{1}{4}$ and $\frac{1}{3}$]. It appears very
computationally challenging to compute separability probabilities for 
values of $q$ other than $\frac{1}{2}$ and 1, although it is an intriguing
hypothesis that they are equal to $\sqrt{2}-1$ for {\it all} (positive) 
$q$.)

We applied the H{\"u}bner formula (\ref{hub2}) for the Bures metric
to this family of $4 \times 4$ density matrices, considering $q$
as a freely-varying parameter, along with $b_{q}$ and $\sigma_{q}^2$.
Computing the $3 \times 3$ 
Bures metric tensor, and {\it then} setting 
$q=1$, we obtain the metric
\begin{equation} \label{AbeRajextended}
ds_{AbeRaj_{q=1}}(\rho,\rho+ d\rho)^2  = \frac{c}{1024} dq^2 +
\end{equation}
\begin{displaymath}
\frac{\log (-2\,{\sqrt{2}}\,{b_q} + {{{\sigma }_q}}^2) - 
    \log (2\,{\sqrt{2}}\,{b_q} + {{{\sigma }_q}}^2)}{8\,
    {\sqrt{2}}} dq db_{q} + 
\end{displaymath}
\begin{displaymath}\frac{2\,\log (8 - {{{\sigma }_q}}^2) - 
    \log (-2\,{\sqrt{2}}\,{b_q} + {{{\sigma }_q}}^2) - 
    \log (2\,{\sqrt{2}}\,{b_q} + {{{\sigma }_q}}^2)}{32} dq d\sigma_{q}^2 +
\end{displaymath}
\begin{displaymath}
\frac{\sigma_{q}^2}{-32 b_{q}^2 +4 (\sigma_{q}^2)^2} (db_{q})^2 +\frac{b_{q}}{16 b_{q}^2 -2 (\sigma_{q}^2)^2 } d b_{q} d \sigma_{q}^2 +\frac{b_{q}^2 -\sigma_{q}^2}{4 (-8 +\sigma_{q}^2) (-8 b_{q}^2 +(\sigma_{q}^2)^2)} (d \sigma_{q}^2)^2.
\end{displaymath}
Here, 
we have 
\begin{equation}
c = -4\,{\log (8 - {{{\sigma }_q}}^2)}^2\,{{{\sigma }_q}}^2\,
   \left( -8 + {{{\sigma }_q}}^2 \right)  + 
  2\,\log (-2\,{\sqrt{2}}\,{b_q} + {{{\sigma }_q}}^2)\,
   \log (2\,{\sqrt{2}}\,{b_q} + {{{\sigma }_q}}^2)\,
   \left( 8\,{{b_q}}^2 - 
     {{{{\sigma }}_q}}^4 \right)
\end{equation}
\begin{displaymath}
  - 
  {\log (-2\,{\sqrt{2}}\,{b_q} + {{{\sigma }_q}}^2)}^2\,
   \left( 8\,{{b_q}}^2 + 
     {{{\sigma }_q}}^2\,
      \left( -16 + {{{\sigma }_q}}^2 \right)  - 
     4\,{\sqrt{2}}\,{b_q}\,
      \left( -8 + {{{\sigma }_q}}^2 \right)  \right) 
\end{displaymath}
\begin{displaymath}
 - 
  {\log (2\,{\sqrt{2}}\,{b_q} + {{{\sigma }_q}}^2)}^2\,
   \left( 8\,{{b_q}}^2 + 
     {{{\sigma }_q}}^2\,
      \left( -16 + {{{\sigma }_q}}^2 \right)  + 
     4\,{\sqrt{2}}\,{b_q}\,
      \left( -8 + {{{\sigma }_q}}^2 \right)  \right) +
\end{displaymath}
\begin{displaymath}
  4\,\log (8 - {{{\sigma }_q}}^2)\,
   \left( -8 + {{{{\sigma }}_q}}^2 \right) \,
   \left( \log (-2\,{\sqrt{2}}\,{b_q} + 
        {{{\sigma }_q}}^2)\,
      \left( -2\,{\sqrt{2}}\,{b_q} + 
        {{{{\sigma }}_q}}^2 \right)  + 
     \log (2\,{\sqrt{2}}\,{b_q} + {{{\sigma }_q}}^2)\,
      \left( 2\,{\sqrt{2}}\,{b_q} + 
        {{{{\sigma }}_q}}^2 \right)  \right).
\end{displaymath}

Numerical computations indicate that the volume element of the metric 
$ds_{AbeRaj_{q}}(\rho,\rho+d \rho)^2$, for any value of $q$, is zero.
So, we have, to this point in our analyses, yet to find  
a {\it nondegenerate}
$q$-extension of the Bures metric (if one is so possible).
(We investigated the possibility of analyzing the $4 \times 4$ density matrix
in \cite[eq. (14)]{tlb}, but it appears to have one zero eigenvalue, thus 
rendering the  H{\"u}bner formula (\ref{hub2}) inapplicable (cf. \cite[sec. 3.4]{hans1}).)

In the {\it unextended} (two-parameter) case, 
the {\it nondegenerate} volume element (with $q=1$) is 
\begin{equation}
V_{AbeRaj_{q=1}}= \frac{{\sqrt{-\left( \frac{1}
         {\left( -8 + \sigma_{q} \right) \,
           \left( -8\,b_{q}^2 + (\sigma_{q}^2)^2 \right) } \right) }}}{4}.
\end{equation}

\section{Concluding Remarks}
Naudts \cite{naudts} 
introduced the concept of a $\phi$-exponential family of 
density operators $\rho_{\theta}$ (for which the obvious example is $\phi(u)=u^q$).
He showed that the $\phi$-exponential family of density operators, together
with a family of escort density operators, optimizes a generalized version
of the well-known Cram\'er-Rao lower bound. He assumes that certain
Hamiltonians are two-by-two commuting. Therefore, the quantum information
manifold $(\rho_{\theta})_{\theta}$ is abelian, which ``is clearly too
restrictive for a fully quantum-mechanical theory''. He suggests further work to remove this restriction.

Abe regarded the order of the escort distribution $q$ as a parameter \cite{abe1}. He
studied the geometric structure of the one-parameter family of
escort distributions using the Kullback divergence, and showed that the
Fisher metric is given in terms of the generalized bit variance, which
measures fluctuations of the crowding index of a multifractal.

\begin{acknowledgments}
I wish to express gratitude to the Kavli Institute for Theoretical
Physics (KITP)
for computational support in this research.

\end{acknowledgments}

\bibliography{BadBures}

\end{document}